# Convergence of Corporate and Information Security

Syed (Shawon) M. Rahman, PhD
Assistant Professor of Computer Science
University of Hawaii-Hilo, HI, USA and
Adjunct Faculty, Capella University, MN, USA
.

Shannon E. Donahue   CISM,CISSP
Ph. D. Student, Capella University,
225 South 6th Street, 9th Floor
Minneapolis, MN 55402, USA
.

*Abstract*—As physical and information security boundaries have become increasingly blurry many organizations are experiencing challenges with how to effectively and efficiently manage security within the corporate. There is no current standard or best practice offered by the security community regarding convergence; however many organizations such as the Alliance for Enterprise Security Risk Management (AESRM) offer some excellent suggestions for integrating a converged security program. This paper reports on how organizations have traditionally managed asset protection, why that is changing and how to establish convergence to optimize security's value to the business within an enterprise.

## I. INTRODUCTION

Throughout history organizations have had security and loss prevention departments to protect their physical assets. In the last 10 to 20 years however there has been a major shift in what is considered an asset. Information and intangible assets have increased significantly. In fact, one could make a very strong argument that information has become an enterprise's most important asset. In the not so distant past, organizations considered their most valuable assets to be physical assets. With the growth of the internet and increased methods for communication which has given most organizations the ability to do business globally, along with the expansion of data warehousing and electronic storage, information has fast become the most critical element to the success of an organization [1].

Information provides organizations with data on their customers, finances, inventories, suppliers, partners and competitors. Metadata (data formed by combining groups of information) provides organizations with vital information that organizations use for decision making on a daily basis. Even the organizations who do still mainly rely on physical resources need information to forecast and communicate with vendors and suppliers.

Before intangible assets became the largest value to organizations throughout the world, most companies counted their physical assets as their primary asset. Organizations had loss prevention units and security departments that safeguarded the company's assets with cameras, physical access control measures, and security guards.

These corporate security departments were made up largely of former law enforcement officers that reported to legal, compliance or risk management divisions. The main charges of this department were intrusion protection and investigations in the workplace. As the digital age came to life, organizations saw more and more information being stored on servers, in databases, and in files.

Organizations began to realize the value that intangible assets provide and subsequently created departments responsible for securing that information. IT security departments and information security departments were tasked with managing the risk that surrounded information.

These information security departments were created after organizations saw threats to their information in the form of hackers, malware, unavailability, and data theft. Additionally, both regional and international laws have surfaced which require dedicated information security managers to be responsible for a formal information security program which is responsible for data protection. Many of the people responsible for information security were moved into the role due in part to their technical backgrounds which were very useful when considering technical controls and designs of a logical perimeter. However, due to their technical backgrounds most of these information security professionals had no experience in traditional corporate or physical security.

Initially having two separate groups responsible for the security of different types of assets was not a problem. However, some security functions have begun to converge on their own and this creates unique challenges for organizations with stove piped security functions.

As technology evolved corporate security departments began to use some advanced tools and technology such as traditional closed circuit cameras running over IP networks. The cameras were the responsibility of the corporate security team, but IT had control of the IP network. The same problem happened with access cards. The corporate security department traditionally had control over physical access control but with databases housing all of the data, IT and information security again were clearly involved. Vendors have recognized this trend and have began to offer converged solutions for risks such as access control [12].

It occurred to many that having the corporate and information security departments work together may be able help to reduce an organization's overall risk profile while streamlining some redundant security processes. However, convergence did not look so appealing to everyone. While the unintentional convergence of systems was happening, the two departments who were initially affected were not working together to improve the risk.

When discussing information and corporate security, convergence is defined as "a trend affecting global enterprises that involves the identification of risks and interdependencies between business functions and processes within the enterprise and the development of managed business process solutions to address those risks and interdependencies" [1]. This definition speaks to the need for organizations to look at convergence and begin to disassemble the organizational silos in order to encourage collaboration to manage risk holistically. Fostering an environment rich in collaboration will help the organization lower its overall risk and reach its business objectives.

According to a 2007 article in security magazine [15], "the silo approach to managing enterprise risks is inadequate because it leaves too many gaps and provides no reliable way to evaluate an enterprise's risk position" [15]. Due in no small part to these gaps, many organizations have tried to begin to explore converging the functions. In reality, convergence between information and corporate security is still very immature. In most organizations, even if the departments both report to a single Chief Security Officer (CSO) or Chief Risk Officer (CRO) the departments are still as different as apples and oranges acting in their traditional silos and worse than that oftentimes shutting one another out.

## II. REASONS FOR CONVERGENCE

One reason for convergence is that threats continually increase and become threats to both corporate and information security safeguards. If an organization's corporate and information security departments do not work together they may miss out on valuable information that could be beneficial to both areas regarding particular risks and threats. Additionally, working in silos may be detrimental as downstream risks that are an emerged result of other risks may not be considered.

United States Computer Emergency Readiness Team (US-CERT)[2] interacts with federal agencies, industry, the research community, state and local governments, and others to collect reasoned and actionable cyber security information and to identify emerging cyber security threats. Based on the information reported, US-CERT was able to identify the following cyber security trends (figure 2 and 3) for fiscal year 2009 first quarter [2].

Figure 1 displays the overall distribution of cyber security incidents and events across the six major categories. The percentage of Category 5 (Scans, Probes, or Attempted Access) reports decreased for the second consecutive quarter. This was a 2.9% decrease in CAT 5 incidents compared to the previous quarter. The percentage of Malicious Code incidents increased by 3.3%. Figure 2 is a breakdown of the top five incidents and events versus all others. Phishing remained the most prevalent incident type, accounting for 70% of all incidents reported. This was a slight percentage decrease of 1.8% from the previous quarter. on the other hand, The sophistication of attack tools has gone up, while the level of skill required to use those tools has gone down(refer to figure 3). At this stage, the attacker takes advantage of his or her ability to steal confidential and proprietary data and sells it for profit or uses it for military intelligence [3].

Another strong argument for convergence is the blurring of boundaries between corporate and information security. For example, if a corporate security department is in charge of corporate access control to restricted areas via card readers, but IT owns the systems who responds when there is a major breach? Likewise, if there is a disaster and information security and corporate security are not aligned with their plans whose plan does the organization follow ?

Convergence offers the organization the opportunity to restructure systems. Currently, systems in physical and information security are oftentimes segregated from one another and are not aware of what the other systems are doing. Once an organization decides to move forward with convergence, systems can be combined which saves the organization money on upkeep and maintenance of infrastructure, not having to purchase any new hardware and also lowers bandwidth uses on the company network.

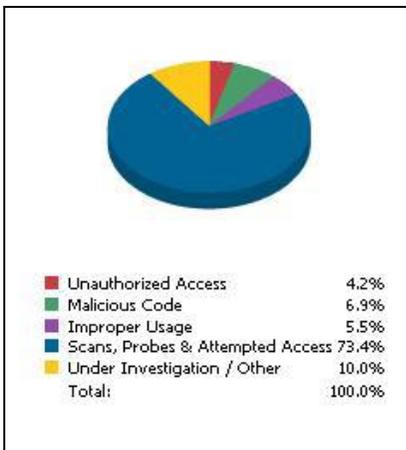

Figure 1: Incidents and Events by Category[2]

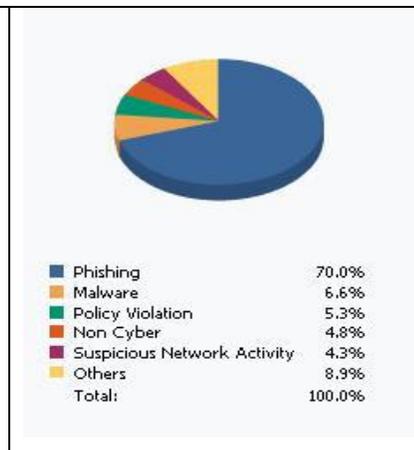

Figure 2: Top Five Incidents vs. All Others [2]

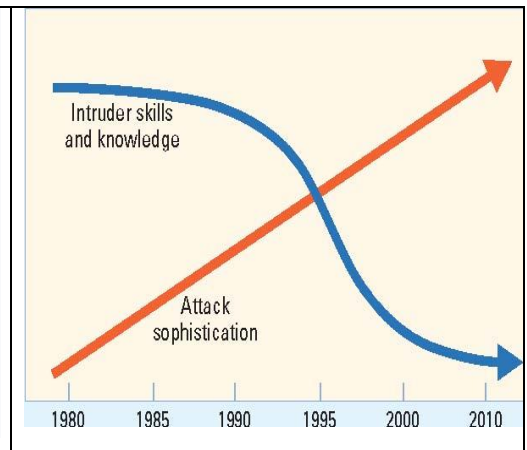

Figure 3. Attack tool trends[3]

### III. BENEFITS OF CONVERGENCE

Joining the information and corporate security departments in some way will help to dispel some of the device management confusion. Whether it is through just working together to have information security and corporate security handle a joint presentation to promote awareness training, or actually joining the two separate functions to report to one CRO or one CSO benefits can be derived from some sort of convergence. If organizations choose to organizationally link the two functions there will be some immediate cost savings as a Chief managing both areas should have an understanding of which team members should be managing particular areas and can set goals that are achievable by both departments. This helps to enable the departments to work together towards a common goal and hopefully will reduce the amount of overall risk to the enterprise.

Gaps between corporate and information security have caused problems in the past and convergence including collaboration and training can help to minimize these gaps. For example, if the information security department has all controls applied appropriately and theft occurs as a result of a thief posing as an employee the breach has still occurred. In fact, a good example of information and corporate security not being on the same page occurred recently at the Sumitomo Mitsui Bank in London, England. Criminals who posed as janitors within the bank had installed devices on computer keyboards that allowed them to obtain login information. The criminals tried unsuccessfully to steal £220 million. Information security controls had been applied but there was a physical security breach that could have been devastating to the organization [6].

While convergence could surely have helped in the above mentioned situation, many organizations are slow to adopt a converged approach. CSO's and CISO's have spoken up and want the business community to understand the benefits of convergence. It appears that business leaders are becoming more accepting of the idea of convergence. According to a 2005 PricewaterhouseCoopers and *CIO* magazine survey of 8,200 IT and security executives in 63 countries of, 53% of organizations have some level of integration between their corporate and IT security divisions [8]. That's up from just 29% in 2003. The projections seem to be growing as well. The Alliance for Enterprise Security Risk Management expected 2005 global spending on convergence activities expected to reach $1.1 billion dollars and significantly move upward after that.

Some of the major benefits that organizations are seeing as a result of converged information and corporate security are significant. Many organizations are saving millions of dollars by streamlining these functions.

Other benefits exist as well. It is hard to put a dollar amount on how safe people feel in the office, or how not being the latest company to lose millions of customer's data is effecting the organization. But through these types if situations, security is not only keeping its assets safe but is preventing unnecessary funds from going out the door as well.

According to a computerweekly.com report executives are seeing the benefits of convergence to organizational risk, "According to the results of a global survey conducted by the Economist Intelligence Unit (EIU) for AT&T, the majority of executives (52%) believe that having a converged network gives their companies a better defense against IT security breaches" [11].

TABLE 1: GLOBAL SECURITY CONVERGENCE SPENDING FORECAST [4].

| | 2004 | 2005 | 2006 | 2007 | 2008 |
|---|---|---|---|---|---|
| Large-scale convergence projects in NA and Europe | 19 | 68 | 175 | 382 | 856 |
| Physical/logical access control projects in NA and Europe | 50 | 150 | 413 | 903 | 1,656 |
| Other projects performed jointly by IT and physical security departments in NA and Europe | 13 | 45 | 118 | 246 | 406 |
| Public sector: border control convergence systems, law enforcement projects in NA and Europe | 410 | 820 | 1,899 | 4,202 | 8,003 |
| Small projects (data center security, communications security, etc.) in NA and Europe | 14 | 40 | 108 | 229 | 369 |
| Total | 506 | 1,123 | 2,713 | 5,962 | 11,289 |

Cost control and productivity may also be improved as a result of convergence. By having systems joined, an organization can eliminate steps that are redundant. This helps to improve processes, eliminate human errors, and increase productivity which ultimately generates revenue for the organization.

Organizations have learned quite a bit about business continuity and disaster recovery in the last few years. September 11, Hurricane Katrina, and devastating natural disasters around the world have brought an increased focus on business continuity (BCP) and disaster recovery planning (DRP). BCP and DRP are another area where convergence plays an important role. We believe information technology security and corporate security must work together to ensure that they are meeting one another's business needs when it comes to project their assets from inside and outside threats and recover data and resources from any attacks.

The collaboration will also streamline efforts so that everyone can understand what to do in cases of emergencies. Collaborating on these efforts will be instrumental in restoring

4*(IJCSIS) International Journal of Computer Science and Information Security,*
*Vol. 7, No. 1, 2010*services in a timely manner. Restoration of these services is an extremely critical issue and is an indirect revenue generator since the quicker that services are restored the faster money is being made.

While the benefits of convergence seem very obvious, there are quite a few challenges that will create some issues for the person who is charged with convergence.

## IV. CHALLENGES OF CONVERGENCE

Security convergence is not without its challenges. The information security department and corporate security departments have long operated in parallel. They have not shared and have not wanted to and, in a lot of situations have through of the other as an adversary. In order to have convergence be effective, collaboration between these two areas must happen. It is not enough to have them reporting through the same channel, they must both have common goals and have mutual discussions on how to meet those goals. In order to get to this level of cooperation there are huge cultural challenges that will need attention.

Information security professionals and corporate security professionals do not usually have the same background. Many information security professionals have technical backgrounds while many of the corporate security professionals have law enforcement backgrounds. According to Steve Hunt, President of 4A international LLC a security consulting firm in Chicago, these differences can lead to a gap in how the two departments evaluate security technologies and controls [8].

Salaries are another issue. Corporate security professionals are not earning comparable salaries generated by information security professionals. Information security professionals often collect six figure salaries early in their career whereas the traditional corporate security professional might be making only half of that closer to the end of their career. Parity adjustments are not practical because there are ways to justify the salary disparity such as technical skills, higher levels of education, professional certifications, and a better job market. However, when an organization combines groups and has them working in tandem, if half of them are earning 50% of what the others are earning there will be some bitter feelings which may result in a hostile environment where people will not work together.

One additional challenge is the training gap. Corporate security professionals often have not been trained on information security or technology. They are not aware of what to watch for when performing their own duties. Often they are unhappy with the convergence plan because they are concerned that they will not retain their positions. A cross training plan that would help the corporate security people understand better what social engineering is and how to spot it, what phishing is and how to spot it and some other technical training would go a long way towards team building and improved skill sets. Additionally, information security professionals likely have very little skill in surveillance, disaster response, investigations and loss prevention. Both sets of individuals need to have some basic training so they are at least aware of what the other is doing and how it complements their own security efforts.

Organizational culture is the most difficult thing to influence so any help from senior management is going to have added benefit. It is important that senior management really demonstrate support for the security program so that employees understand that it is a critical piece of the business that is everyone's responsibility in the same way that customer service and quality management are.

TABLE 2: SUMMARY OF BENEFITS AND CHALLENGES OF CONVERGENCE

| Benefits and challenges of Convergence ||
|---|---|
| **Benefits** | **Challenges** |
| Cost Savings | Culture |
| More holistic view of risk | Salary Differences |
| Reduction of risk profile | Training requirements |
| Streamline process | Lack of Collaboration |

## V. BEGINNING A CONVERGED PROGRAM

Once the organization decides that they want to converge their security program there are many things to consider. Organizational structure is one, budgets are another and the overall risk profile is a third.

### A. Organizational Structure

Organizationally there are many routes that can be taken to align the physical and information security organizations. Each of the options has positives and negatives and may work in one organization and not at all in another. Security managers need to know the organization that they work in, consider their budgeting options as well as objectives and select a solution that will meet the needs of their business.

The first method is to combine the physical and information security departments into one security organization reporting to a CSO or CRO. This group would have responsibility for all things security from the guard at the front desk to encryption protocols for sensitive data.

While the model of combining all the players does force some level of integration, and if successful can be extremely effective in business process integration and fraud detection it is not without drawbacks. Going right to a fully immersed security program will increase the likelihood that the security staff will be upset. This unhappiness can result in a lack of cooperation and hostile work environments and may even cause damage to the risk profile because staff is so worried about their work situation that they aren't paying attention to the organizational risks.

If this is the route chosen the security manager will need to spend plenty of time working on opening communication channels and fostering collaboration. Cross training will be an immediate need if expectations are for people to learn each

66                                    http://sites.google.com/site/ijcsis/
ISSN 1947-5500

other job functions and additional funding for corporate training may be necessary depending on the gap in skills.

The next option is to keep the physical and logical departments as separate departments with their own budgets and reporting lines but ultimately have them report to the same executive, most likely a CSO. The CSO would receive information from both security areas and be able to make decisions based on the information provided to them by the business unit director. The groups would most likely not work joint projects or go through any cross training, but would still benefit from updated processes based on organizational goals. The departments would still need to collaborate to ensure all risks are addressed. While this level of integration cannot be considered complete immersion, security does not have to be a completely combined area to reap some of the benefits of convergence. This type of management can still be extremely effective at driving down costs and eliminating redundant work and systems.

The third option which is becoming more popular these days is to keep the functions completely separate and assist in process management and collaboration through bringing security issues to a risk council staffed with business and security management that could make decisions regarding security. This approach would be helpful in minimizing culture issues and would probably be well received by members of both security teams.

Whichever organizational structure choice is taken it is important to consider the culture of the organization. Culture is an often overlooked area that can be critical to the success of implementing convergence and should not be underestimated.

Many things can be done to smooth the progress of convergence. Combining processes where possible, gaining senior management support for the initiative and beginning to look at organizational risks instead of just risks to individual departments will help to incorporate convergence smoothly.

Many frameworks exist to help an organization lower their risk profile. It is important to remember that the reason for discussing convergence of information security and physical security is to show how security professionals can help to improve the overall risk profile in your organization.

Some corporate risk management frameworks that are used internationally are the Committee of Sponsoring Organizations' (COSO) Enterprise Risk Management (ERM) framework, the Operationally Critical threats, Asset and Vulnerability Evaluation model (OCTAVE) and one more accepted framework for managing enterprise risk is the Risk and Insurance Management Society (RIMS) Risk Management Model (RMM). All of these models focus on managing operational risk and reducing the overall risk profile.

We have found the AESRM recommendation useful while beginning a converged program [4]:

- ❖ Establish a governance framework for managing security risks.
- ❖ Define security requirements early in the planning stage.
- ❖ Understand the technology better.
- ❖ Analyze and understand security-related cost-benefit trade-offs.
- ❖ Develop a unified set of meaningful standards.
- ❖ Deploy special network security controls.
- ❖ Implement effective authorization, accountability and auditability controls.
- ❖ Include critical systems in organization continuity plans.
- ❖ Protect information important for investigations.
- ❖ Increase auditing and logging.
- ❖ Require tailored training and awareness programs.
- ❖ Pressure vendors to play a more active role in security.
- ❖ Expand audit coverage of systems and devices.

VI. CONCLUSION

Organizational risks and threats are changing every day. As new technologies expand into the organization the risk profile continues to expand. Many organizations are in a constant state of growth. The organizational risk profile needs to constantly be reviewed and updated.

Since the beginning of information security departments, corporate and information security have operated in their own silos with separate management teams, different risks, different processes, and different budgets. In the last decade, organizations have seen dramatic changes in their risk profile. With technical advances, traditional corporate security functions have evolved to include the use of networks, databases, and file servers. Compliance requirements have dictated how sensitive information needs to be transmitted and so traditional corporate security departments and information security departments find some of their functions that used to have clear management delineations becoming blurry.

Although convergence has been a buzz word for a long time, companies have just recently started to notice the benefits of converging traditional siloed security functions to allow for an enterprise wide view of risk. Some of the benefits that can be derived from converging information and corporate security areas including streamlining processes, saving money on infrastructure, increasing productivity, positioning the organization better to have an broader view of organizational threats and risks and meeting compliance requirements.

Many organizations have attempted to move towards a converged security area and have realized that there are many hurdles to be jumped on the way to a converged organization. One of these challenges is getting the security staff in both areas to work together. Organizations can work to encourage collaboration by ensuring that employee concerns are addressed and people feel confident that communication will be well received. Gaining support from senior management and combining processes to minimize confusion are two more techniques that will prove useful during the implementation of a converged security structure.

While the challenges may be difficult to work through, the benefits that the organization will realize as a result of taking a wider view of risk will be immense. As risks can change at any moment it is absolutely critical to have as many well trained professionals as possible working together to improve the productivity and sustainability of an organization with a holistic approach to security management through convergence in enterprise security risk management

AUTHORS PROFILE

**Syed (Shawon) M. Rahman**

Syed (Shawon) Rahman is an Assistant Professor in the Department of Computer Science & Engineering at the University of Hawaii-Hilo (UHH), Hawaii, USA and an Adjunct Faculty in the Capella University, Minneapolis, USA. He earned his Ph.D. in Software Engineering and MS in Computer Science degree from North Dakota State University. Dr. Rahman's research interests include Information Assurance & Security, Data Visualization, Data Modeling, Web Accessibility, and Software Testing & Quality Assurance. His research focus also includes Software Engineering Education, Search Engine Optimization, and Lightweight Software Development Methodologies such as eXtreme Programming and Test-driven Development.

Before joining at the UHH in Fall 2009, Dr. Rahman taught the last three years in the Dept. of Computer Science and Software Engineering at the University of Wisconsin-Platteville as an Assistant Professor. Dr. Rahman is always interested in applying Emerging Technologies and Tools in classrooms to improve students' learning experience and performance. He has published enormous number referred articles and presented his works around the world. He is an active member of many professional organizations including ACM, ASEE, ASQ, IEEE, and UPE.

**Shannon E. Donahue**

Shannon is a Certified Information Security Manager (CISM) and Certified Information Systems Security Professional (CISSP) with more than 10 years of experience in information technology and security. She was the Information Security Officer at William Rainey Harper College where she was responsible for developing and managing the information security program. She was responsible for securing data on all systems as well as incident management, policy development, and security awareness.

As a network support analyst and security architect for AT&T, Shannon provided tier 3 network support and helped to develop the virus response plan for managed service customers. She also was a lead member responsible for incident response and disaster recovery plans. In her current position with ISACA, Shannon is responsible for managing the security program and for serving the needs of the security profession through research projects and publications. Shannon has a masters degree in Management & Systems from New York University and is currently working on a PhD in Information Security at Capella University.